\documentclass[showpacs,twocolumn]{revtex4}
\usepackage[T1]{fontenc}
\usepackage[latin1]{inputenc}
\usepackage{amsmath}
\usepackage{amssymb}
\usepackage{dsfont}
\usepackage{color}
\usepackage{multirow}
\makeatletter
\usepackage{hyperref}
\usepackage{amssymb}
\usepackage{graphicx}
\usepackage{epstopdf}
\usepackage{epsfig}
\usepackage{subfigure}
\usepackage{dsfont}
\usepackage{tikz}
\makeatother
\usepackage{float}
\restylefloat{table}

\renewcommand{\vec}[1]{\mathbf{#1}}

\usepackage{cancel,ifthen}
\usepackage[normalem]{ulem}

\newcommand{\comment}[2][NoInPuT]{\ifthenelse{\equal{#1}{NoInPuT}}{}{{\color{blue}\sout{#1}}}{\color{red} #2}}
\begin{document}

\title{Kinetic Model of Trapped Finite Temperature Binary Condensates}

\author{M. J. Edmonds}
\author{K. L. Lee}
\author{N. P. Proukakis}
\affiliation{Joint Quantum Centre (JQC) Durham-Newcastle, School of Mathematics and Statistics,
Newcastle University, Newcastle upon Tyne NE1 7RU, England, UK}

\pacs{03.75.Mn,67.85.-d}
\date{\today{}}

\begin{abstract}\noindent
We construct a non-equilibrium theory for the dynamics of two interacting finite-temperature atomic Bose-Einstein condensates and use it to numerically estimate the relative rates of the arising collisional processes near equilbrium. The condensates are described by dissipative Gross-Pitaevskii equations, coupled to quantum Boltzmann equations for the thermal atoms. The density-density interactions between atoms in different components facilitate a number of transport processes of relevance to sympathetic cooling: in particular, 
considering realistic miscible and immiscible trapped atomic $^{87}$Rb-$^{41}$K and $^{87}$Rb-$^{85}$Rb condensate mixtures, we highlight the dominance of
an inter-component scattering process associated with collisional "exchange" of condensed and thermal atoms between the components 
close to equilibrium.
\end{abstract}
\maketitle
{\it Introduction. }
%
The unprecedented control of trapped neutral cold atom experiments enables the creation and study of degenerate multi-component systems, including Bose-Bose \cite{hall_1998}, Bose-Fermi \cite{hadzibabic_2002} and Fermi-Fermi \cite{demarco_1999,taglieber_2008} mixtures, and the related problems of spinor gases \cite{stamperkurn_2013} and artificial gauge fields \cite{dalibard_2011}. 
In two-component systems, cooling to quantum degeneracy is typically performed through a combination of evaporative \cite{ketterle_1996} and sympathetic \cite{myatt_1997} cooling techniques. 
For bosonic mixtures, this leads to condensation in differing hyperfine states of an atom ($^{87}$Rb~\cite{myatt_1997,hall_1998,maddaloni_2000,mertes_2007}), differing isotopes (e.g. $^{87}$Rb-$^{85}$Rb~\cite{papp_2008}, $^{168}$Yb-$^{174}$Yb~\cite{sugawa_2011}) or differing elements (e.g.~$^{87}$Rb-$^{41}$K~\cite{modugno_2002,thalhammer_2008}, $^{87}$Rb-$^{133}$Cs~\cite{mccarron_2011,lercher_2011}, $^{87}$Rb-$^{84}$Sr and $^{87}$Rb-$^{88}$Sr~\cite{pasquiou_2013}, $^{87}$Rb-$^{23}$Na~\cite{xiong_2013}). 
Interest in such systems has focused on understanding numerous properties, including 
equilibrium profiles \cite{ho_1996,pu_1998,ohberg_1998,trippenbach_2000,pattinson_2013}, stability properties \cite{ohberg_1999,robins_2001}, collective excitations \cite{pu_1998,zhang_2006}, vortices \cite{mueller_2002}, solitary waves \cite{berloff_2005}, 
and dissipative~\cite{pattinson_2014} and quenched dynamics \cite{liu_2014}; 
these effects depend critically on the relative inter-atomic interactions within and between the species, which determine whether the emerging condensates overlap spatially or phase-separate~\cite{ho_1996,pu_1998}.

The description of bosonic mixtures is typically either focused on the low temperature (Gross-Pitaevskii) \cite{pethick_2008}, or high-temperature (Boltzmann) limit \cite{delannoy_2001}, or on treating the condensate in contact with a static heat bath~\cite{lewenstein_1995,timmermans_1998,papenbrock_2002}. Alternative approaches are based on 
classical field methods (which ignore the dynamics of the high-lying thermal modes \cite{bradley_2014,liu_2014}), or on number-conserving methods (which explicitly include only the back-action of the thermal cloud on the condensate \cite{billam_2013,mason_2014}). Moreover, although related kinetic models have been 
derived in the context of spinor gases \cite{nikuni_2003,endo_2011}, there has been to date no critical assessment of the 
{relative importance} of the various collisional processes at finite temperatures.
%

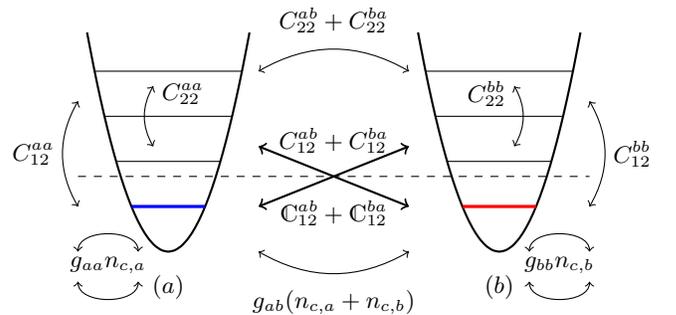
\begin{figure}[b]
\begin{tikzpicture}[scale=0.40]
\draw[color=black,thick,domain=-2.7:2.7,smooth,variable=\x,black] plot ({\x},{\x*\x}); 
\draw[blue,very thick] (-1.22,1.5) -- (1.22,1.5) node[above] {};
\draw [-] (-1.73,3) -- (1.73,3) node[above] {};
\draw[-] (-2.12,4.5) -- (2.12,4.5) node[above] {};
\draw[-] (-2.45,6) -- (2.45,6) node[above] {};
\draw[dashed] (-3,2.5) -- (14,2.5) {};
\draw[color=black,thick,domain=8.3:13.7,smooth,variable=\x,black] plot ({\x},{(\x-11)*(\x-11)});
\draw[red,very thick] (9.78,1.5) -- (12.22,1.5) node[above] {};
\draw[-] (9.26,3) -- (12.73,3) node[above] {};
\draw[-] (8.88,4.5) -- (13.12,4.5) node[below] {};
\draw[-] (8.55,6) -- (13.45,6) node[below] {};
\draw[<->,thick] (3,1.5) -- (8,3.5) node[above] {};
\draw[<->,thick] (3,3.5) -- (8,1.5) node[above] {};
\draw (-3,1.5) edge[<->,black,out=120,in=240] (-3,5);
\draw (14,1.5) edge[<->,black,out=60,in=300] (14,5);
\draw (3,6) edge[<->,black,out=30,in=150] (8,6);
\draw (3,0) edge[<->,black,out=330,in=210] (8,0);
\draw (-0.5,3.5) edge[<->,black,out=120,in=240] (-0.5,5.5);
\draw (11.5,3.5) edge[<->,black,out=60,in=300] (11.5,5.5);
\draw (-3,0) edge[<->,black,out=90,in=90] (-1,0);
\draw (-3,-1) edge[<->,out=270,in=270] (-1,-1);
\draw (12,0) edge[<->,out=90,in=90] (14,0);
\draw (12,-1) edge[<->,out=270,in=270] (14,-1);
\node [black,below] at (0,-0.5) {$(a)$};
\node [black,below] at (11,-0.5) {$(b)$};
\node [black,left] at (-3.5,3.25) {$C^{aa}_{12}$};
\node [black,right] at (14.5,3.25) {$C^{bb}_{12}$};
\node [black,above] at (5.5,7) {$C^{ab}_{22}+C^{ba}_{22}$};
\node [black,below] at (5.5,-1) {$g_{ab}(n_{c,a}+n_{c,b})$};
\node [black,above] at (5.5,3) {$C^{ab}_{12}+C^{ba}_{12}$};
\node [black,below] at (5.5,2) {$\mathds{C}^{ab}_{12}+\mathds{C}^{ba}_{12}$};
\node [black,right] at (-0.5,5.25) {$C^{aa}_{22}$};
\node [black,left] at (11.5,5.25) {$C^{bb}_{22}$};
\node [black,above] at (-2,-1) {$g_{aa}n_{c,a}$};
\node [black,above] at (13,-1) {$g_{bb}n_{c,b}$};
\end{tikzpicture}
\centering
\caption{(Color online) Schematic representation of collisional model. The two trapped condensates are denoted $(a)$ (left) and $(b)$ (right). Arrows indicate the various mean field and collisional transport processes occurring within and between condensates and non-condensates of different components.}
\label{fig1}
\end{figure}

{In this work we (i) present a self-consistent model for the study of partly-condensed bosonic mixtures in the presence of dynamical condensates and thermal clouds (shown schematically in Fig.~1),
and (ii) perform detailed numerical calculations at equilibrium to shed light into the relative importance of those processes for experimentally-accessible miscible and immiscible mixtures. 
A key conclusion is the estimation of the importance of a novel "cross-condensate-exchange" term found to dominate as soon as both species exhibit condensation.}
%
%


Our approach is a generalization of the "Zaremba--Nikuni--Griffin" (ZNG) kinetic model \cite{zaremba_1999,nikuni_1999,jackson_2002b} which, in the context of single-component condensates, has proven extremely useful in describing collective modes \cite{jackson_2001_2002_2003}, condensate growth \cite{bijlsma_2000}, dynamics of macroscopic excitations \cite{jackson_2007} and surface evaporative cooling \cite{markle_2014}. 
This approach is based on the decomposition of the Bose field operator into a symmetry-breaking part and its fluctuations, and a separation of timescales argument (see also \cite{proukakis_1998,proukakis_2008,walser_1999}): in 
our analysis, we explicitly separate slowly-evolving variables for the condensate and thermal clouds, whose non-vanishing mean values are defined by appropriate dynamical equations, from those evolving on the more rapid  collisional timescales. A careful self-consistent perturbative 
treatment of the latter in the context of an appropriate perturbing Hamiltonian (details to appear elsewhere) 
enables us to quantify for the first time collisional processes not accounted for in earlier models of Refs.~\cite{nikuni_2003,endo_2011}. 

{\it Coupled Kinetic Theory. } The second-quantized Hamiltonian describing the interacting binary mixture is 
{\begin{equation}\label{h1}
\hat{H}=\sum_{j=a}^{b}\int d{\bf r}\ \hat{\Psi}^{\dagger}_{j}\bigg(\hat{H}_{0,j}+\frac{1}{2}\sum_{k=a}^{b}g_{kj}\hat{\Psi}^{\dagger}_{k}\hat{\Psi}_{k}\bigg)\hat{\Psi}_{j},
\end{equation}
where $\hat{\Psi}_{j}\equiv\hat{\Psi}_{j}({\bf r})$ is the annihilation operator for a species-$j$ atom, $\hat{H}_{0,j}=-({\hbar^2}/{2m_j})\nabla^2+V_{j}({\bf r})$} and $g_{kj}=2\pi\hbar^2a_{kj}/m_{kj}$, where $a_{kj}$ is the scattering length between atoms in component $j$ and $k$, $m_{kj}^{-1}=m^{-1}_{k}+m^{-1}_{j}$ is the reduced mass, $m_{j}$ is the mass of a boson in component $j$; and $V_{j}({\bf r})$ is the trapping potential.

The condensed and non-condensed components are separated by the Beliaev decomposition $\hat{\Psi}_{j}=\phi_{j}+\hat{\delta}_{j}$, with condensate evolution equations obtained from the Heisenberg equations of motion for $\hat{\Psi}_{j}$ as
{\begin{align}\label{eom1}\nonumber
i\hbar\frac{\partial\phi_j}{\partial t}=&\bigg[-\frac{\hbar^2}{2m_{j}}\nabla^2+U^{j}_{c}\bigg]\phi_{j}+g_{jj}\bigg[\langle\hat{\delta}_{j}\hat{\delta}_{j}\rangle\phi^{*}_{j}+\langle\hat{\delta}^{\dagger}_{j}\hat{\delta}_{j}\hat{\delta}_{j}\rangle\bigg]\\&+g_{kj}\bigg[\langle\hat{\delta}^{\dagger}_{k}\hat{\delta}_{j}\rangle\phi_{k}+\langle\hat{\delta}_{k}\hat{\delta}_{j}\rangle\phi^{*}_{k}+\langle\hat{\delta}^{\dagger}_{k}\hat{\delta}_{k}\hat{\delta}_{j}\rangle\bigg],
\end{align}
where $\langle \cdots \rangle$ denotes averages (with $\langle \hat{\delta}^{(\dag)}_{j} \rangle = 0$),
$j,k\in\{a,b\}$, and 
$U^{j}_{c}({\bf r},t)=V_{j}({\bf r})+g_{jj}(n_{c,j}+2\tilde{n}_{j})+g_{kj}(n_{c,k}+\tilde{n}_{k})$
is the effective condensate potential.
Following established techniques \cite{zaremba_1999}, 
we identify the condensate fields $\phi_{j}$ (corresponding condensate densities $n_{c,j}=|\phi_{j}|^2$),
and {\em diagonal} non-condensate densities $\tilde{n}_{j}=\langle\hat{\delta}^{\dagger}_{j}\hat{\delta}_{j}\rangle$ as the only relevant {\em slowly-varying} quantities of interest. 
Triplet anomalous averages $\langle\hat{\delta}_{j}^{\dagger}\hat{\delta}_{j}\hat{\delta}_{j}\rangle$, $\langle\hat{\delta}_{k}^{\dagger}\hat{\delta}_{k}\hat{\delta}_{j}\rangle$ are treated perturbatively (via adiabatic elimination \cite{proukakis_1998,proukakis_2008}), giving rise to well-known source terms, and we only maintain energy-conserving contributions, thus also discarding pair anomalous averages $\langle\hat{\delta}_{j}\hat{\delta}_{j}\rangle$, $\langle\hat{\delta}_{k}\hat{\delta}_{j}\rangle$ \cite{griffin_1996}.
An important aspect of our work (see also Refs.~\cite{bradley_2014,mason_2014}) is that we explicitly maintain perturbative corrections arising from the {\em off-diagonal} terms $\langle\hat{\delta}_{k}^{\dagger}\hat{\delta}_{j}\rangle$. As a result,
%
%
the equation of motion for component $j$ takes the form
\begin{equation}
i\hbar\frac{\partial\phi_{j}}{\partial t}=\bigg[-\frac{\hbar^2}{2m_j}\nabla^2+U^{j}_{c}-i(R^{jj}+R^{kj}+\mathds{R}^{kj})\bigg]\phi_j.\label{eom2}
\end{equation}}
{It contains a number of source terms:
$R^{jj}({\bf r},t)=-ig_{jj}\langle\hat{\delta}^{\dagger}_{j}\hat{\delta}_j\hat{\delta}_j\rangle/\phi_j$ 
appearing in Eq.~\eqref{eom2}
describes the intra-component scattering of condensate and non-condensate particles 
and is well-known from single-component ZNG theory 
\cite{zaremba_1999};
$R^{kj}({\bf r},t)=-ig_{kj}\langle\hat{\delta}^{\dagger}_{k}\hat{\delta}_{k}\hat{\delta}_{j}\rangle/\phi_{j}$
is a direct generalization of $R^{jj}$ to binary condensates,
describing the inter-component scattering of condensed and non-condensed atoms.
%
$\mathds{R}^{kj}({\bf r},t)=-ig_{kj}\langle\hat{\delta}^{\dagger}_{k}\hat{\delta}_{j}\rangle\phi_{k}/\phi_{j}$, is qualitatively different from the other two (see Eq. \eqref{c12jk2} below). $R^{kj}$, $R^{jj}$ and $\mathds{R}^{kj}$ are defined through their relation to the collision integrals $C^{kj}_{12}$ ($C^{jj}_{12}$) and $\mathds{C}^{kj}_{12}$ given below.}
\begin{figure}[t]
\centering
\begin{tikzpicture}[scale=0.55]
\usetikzlibrary{arrows}
\usetikzlibrary{patterns}
\tikzset{axes/.style={thick,->},
rr/.style={rectangle,draw,fill=red!100,minimum size=8},
bc/.style={circle,draw,fill=blue!100,minimum size=2},
br/.style={rectangle,draw,fill=blue!100,minimum size=8},
rc/.style={circle,draw,fill=red!100,minimum size=2},}
\node (a) at (0,0) [bc] {};
\node [black,below] at (0,-0.3) {$p_3$};
\node (b) at (1.5,0) [br] {};
\node [black,below] at (0,2.0) {$p_4$};
\node (c) at (0,1) [bc] {};
\node [black,below] at (1.5,-0.3) {$p$};
\node (d) at (1.5,1) [bc] {};
\node [black,below] at (1.5,2.0) {$p_2$};
\draw [dashed] (c.east) -- (b.west);
\draw [dashed] (a.east) -- (d.west);
\node (a) at (2.5,0) [br] {};
\node [black,below] at (2.5,-0.3) {$p$};
\node (b) at (4,0) [bc] {};
\node [black,below] at (4,-0.3) {$p_3$};
\node (c) at (2.5,1) [bc] {};
\node [black,below] at (2.5,2) {$p_2$};
\node (d) at (4.0,1) [bc] {};
\node [black,below] at (4,2) {$p_4$};
\draw [dashed] (c.east) -- (b.west);
\draw [dashed] (a.east) -- (d.west);
\node [black,below] at (2,0.9) {$-$};
\node (a) at (0,-2.5) [bc] {};
\node (b) at (1.5,-2.5) [rr] {};
\node (c) at (0,-1.5) [rc] {};
\node (d) at (1.5,-1.5) [bc] {};
\draw [dashed] (c.east) -- (b.west);
\draw [dashed] (a.east) -- (d.west);
\node (a) at (2.5,-2.5) [rr] {};
\node (b) at (4,-2.5) [bc] {};
\node (c) at (2.5,-1.5) [bc] {};
\node (d) at (4,-1.5) [rc] {};
\draw [dashed] (c.east) -- (b.west);
\draw [dashed] (a.east) -- (d.west);
\node [black,below] at (2,-4.1) {$-$};
\node (a) at (0,-5) [rc] {};
\node (b) at (1.5,-5) [br] {};
\node (c) at (0,-4) [bc] {};
\node (d) at (1.5,-4) [rc] {};
\draw [dashed] (c.east) -- (b.west);
\draw [dashed] (a.east) -- (d.west);
\node (a) at (2.5,-5) [br] {};
\node (b) at (4,-5) [rc] {};
\node (c) at (2.5,-4) [rc] {};
\node (d) at (4,-4) [bc] {};
\draw [dashed] (c.east) -- (b.west);
\draw [dashed] (a.east) -- (d.west);
\node [black,below] at (2,-1.6) {$-$};
\node (a) at (6.5,0) [rr] {};
\node (b) at (8,0) [br] {};
\node (c) at (6.5,1) [bc] {};
\node (d) at (8,1) [rc] {};
\draw [dashed] (c.east) -- (b.west);
\draw [dashed] (a.east) -- (d.west);
\node (a) at (9,0) [br] {};
\node (b) at (10.5,0) [rr] {};
\node (c) at (9,1) [rc] {};
\node (d) at (10.5,1) [bc] {};
\draw [dashed] (c.east) -- (b.west);
\draw [dashed] (a.east) -- (d.west);
\node [black,below] at (8.5,0.9) {$-$};
\node (a) at (6.5,-2.5) [bc] {};
\node (b) at (8.0,-2.5) [bc] {};
\node (c) at (6.5,-1.5) [bc] {};
\node (d) at (8.0,-1.5) [bc] {};
\draw [dashed] (c.east) -- (b.west);
\draw [dashed] (a.east) -- (d.west);
\node (a) at (9.0,-2.5) [bc] {};
\node (b) at (10.5,-2.5) [bc] {};
\node (c) at (9.0,-1.5) [bc] {};
\node (d) at (10.5,-1.5) [bc] {};
\draw [dashed] (c.east) -- (b.west);
\draw [dashed] (a.east) -- (d.west);
\node [black,below] at (8.5,-1.6) {$-$};
\node (a) at (6.5,-5) [rc] {};
\node (b) at (8.0,-5) [bc] {};
\node (c) at (6.5,-4) [bc] {};
\node (d) at (8.0,-4) [rc] {};
\draw [dashed] (c.east) -- (b.west);
\draw [dashed] (a.east) -- (d.west);
\node (a) at (9.0,-5) [bc] {};
\node (b) at (10.5,-5) [rc] {};
\node (c) at (9.0,-4) [rc] {};
\node (d) at (10.5,-4) [bc] {};
\draw [dashed] (c.east) -- (b.west);
\draw [dashed] (a.east) -- (d.west);
\node [black,below] at (8.5,-4.1) {$-$};
\node [black,below] at (5.5,0.8) {$\mathds{C}^{ab}_{12}$};
\node [black,below] at (-1,0.85) {$C^{aa}_{12}$};
\node [black,below] at (-1,-2) {$C^{ab}_{12}$};
\node [black,below] at (-1,-2.9) {$+$};
\node [black,below] at (-1,-3.6) {$C^{ba}_{12}$};
\node [black,below] at (5.5,-1.6) {$C^{aa}_{22}$};
\node [black,below] at (5.5,-4.1) {$C^{ab}_{22}$};
\node [black,below] at (-2,0.85) {$(i)$};
\node [black,below] at (-2,-2.8) {$(ii)$};
\node [black,below] at (11.5,0.85) {$(iii)$};
\node [black,below] at (11.5,-1.6) {$(iv)$};
\node [black,below] at (11.5,-4.1) {$(v)$};
\end{tikzpicture}
\caption{(Color online) Each diagram represents a kinetic energy and momentum
conserving collision between atoms. Squares and circles represent condensate
and thermal atoms respectively. Component $a$ particles are blue, while $b$ are red.
Diagrams for the b component are obtained by interchanging the two colors in
each square and circle.
}
\label{fig2}
\end{figure}
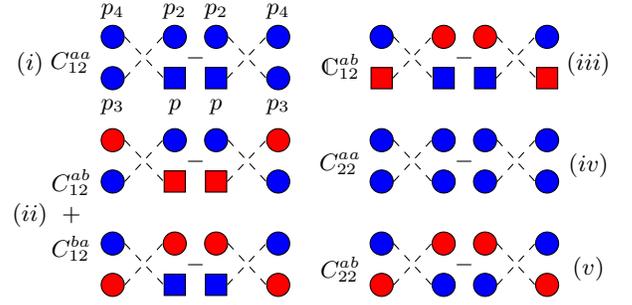

To construct a 
self-consistent theory including the evolution of the non-condensate,
the multi-component single-particle Wigner distribution is introduced as $f^{kj}({\bf p},{\bf r},t)=\int d{\bf r}' e^{i{\bf p}\cdot{\bf r}'/\hbar}\langle\hat{\delta}_{j}^{\dagger}({\bf r}+{\bf r}'/2,t)\hat{\delta}_{k}({\bf r}-{\bf r}'/2,t)\rangle$. We restrict our analysis to the diagonal terms of $f^{kj}({\bf p},{\bf r},t)$, i.e. $f^{jj}({\bf p},{\bf r},t)\equiv f^{j}({\bf p},{\bf r},t)$, valid in the absence of coherent couplings between states. 
%
The kinetic equation for component $j$ is
\begin{align}\nonumber
&\frac{\partial}{\partial t}f^{j}+\frac{1}{m_{j}}{\bf p}\cdot{\nabla_{\bf r}f^{j}}-{\nabla_{\bf p}f^{j}}\cdot{\nabla_{\bf r}U^{j}_{\text{n}}}=\\&\left(C^{jj}_{12}+C^{kj}_{12}\right)+\mathbb{C}^{kj}_{12}+\left(C^{jj}_{22}+C^{kj}_{22}\right).
\label{kin}
\end{align}
The Hartree-Fock potential that non-condensed atoms in component $j$ feel is $U^{j}_{\text{n}}({\bf r},t)=V_{j}({\bf r})+2g_{jj}[n_{c,j}+\tilde{n}_{j}]+g_{kj}[n_{c,k}+\tilde{n}_{k}]$, and the non-condensate density of component $j$ 
is obtained via $\tilde{n}_{j}({\bf r},t)=\int d{\bf p}/(2\pi\hbar)^3 f^{j}({\bf p},{\bf r},t)$. {The source terms appearing in Eq.~\eqref{eom2} are related to the collision integrals by the relationships $-2n_{c,j}R^{kj}/\hbar=\int d{\bf p}/(2\pi\hbar)^3 C^{kj}_{12}$ ($k=j$, $k \neq j$) and $-2n_{c,j}\mathds{R}^{kj}/\hbar=\int d{\bf p}/(2\pi\hbar)^3\mathds{C}^{kj}_{12}$ ($k \neq j$).}
The various collisional contributions to the kinetic scattering of particles for the $a$ component are summarized in Fig.~\ref{fig2}.
Diagrams $(i)$-$(iii)$ concern collision integrals describing the scattering of condensed and non-condensed particles, while diagrams $(iv)$ and $(v)$ show scattering amongst non-condensed atoms. 

The collision integral $C^{kj}_{12}$ (encapsulating $C^{jj}_{12}$) appearing in Eq.~\eqref{kin} is defined by (using $f^{k}_{\mu}\equiv f^{k}({\bf p}_{\mu},{\bf r},t)$)
\begin{align}\nonumber
&C^{kj}_{12}=(1+\delta_{kj})\frac{g_{kj}^{2}}{(2\pi)^2\hbar^4}\ \int d{\bf p}_{2}\int d{\bf p}_{3}\int d{\bf p}_{4}\\
\nonumber&\times \Bigl\{n_{c,k}\,\delta({\bf p}_{c}^{k}+{\bf p}_{2}-{\bf p}_{3}-{\bf p}_{4})\delta(\varepsilon^{k}_{c}+\varepsilon^{j}_{p_2}-\varepsilon^{j}_{p_3}-\varepsilon^{k}_{p_4})\\
\nonumber&\quad\;\,\times[(f^{j}_{2}+1)f^{j}_{3}f^{k}_{4}-f^{j}_{2}(f^{j}_{3}+1)(f^{k}_{4}+1)][\delta({\bf p}-{\bf p}_{2})\\
\nonumber&\quad\;\,-\delta({\bf p}-{\bf p}_{3})]\\
\nonumber&\quad\;\,-n_{c,j}\,\delta({\bf p}_{c}^{j}+{\bf p}_{2}-{\bf p}_{3}-{\bf p}_{4})\delta(\varepsilon^{j}_{c}+\varepsilon^{k}_{p_2}-\varepsilon^{k}_{p_3}-\varepsilon^{j}_{p_4})\\
&\quad\;\,[(f^{k}_{2}+1)f^{k}_{3}f^{j}_{4}-f^{k}_{2}(f^{k}_{3}+1)(f^{j}_{4}+1)]\delta({\bf p}-{\bf p}_4)\Bigr\},
\label{c12jk}
\end{align}
It describes the scattering of a condensate atom and a non-condensate atom into thermal states, and its inverse process (Fig.~\ref{fig2}(i)-(ii)), with the Kronecker delta $\delta_{kj}$  
accounting for enhanced scattering of atoms within the same species ($C^{jj}_{12}$ subcase). 
Here $\varepsilon^{j}_{p}=p^2/2m_j+U^{j}_{\text{n}}$ is the Hartree-Fock energy, $\varepsilon^{j}_{c}=\mu^{j}_{c}+\frac{1}{2}m_{j}v^{2}_{c,j}$ defines the condensate energy, and ${\bf p}^{j}_{c}=m_j{\bf v}_{c,j}$ gives the condensate momentum.

{The collision integral $\mathds{C}^{kj}_{12}$ 
is defined from the average of an {\em off-diagonal pair} of fluctuation operators, $\langle\hat{\delta}^{\dagger}_{k}\hat{\delta}_{j}\rangle$ ($j \neq k$), (in contrast to all other "ZNG" source terms originating from triplet terms $\langle\hat{\delta}^{\dagger}_{k}\hat{\delta}_{k}\hat{\delta}_{j}\rangle$ \cite{nikuni_2003,endo_2011,zaremba_1999}) as}
\begin{align}
\nonumber&\mathds{C}^{kj}_{12}=\frac{2\pi g_{kj}^{2}}{\hbar}\ n_{c,k}\,n_{c,j}\,\int d{\bf p}_{1}\int d{\bf p}_{2}\\\nonumber&\times\delta({\bf p}_{c}^{j}+{\bf p}_{1}-{\bf p}_{c}^{k}-{\bf p}_{2})\delta(\varepsilon^{j}_{c}+\varepsilon^{k}_{p_1}-\varepsilon^{k}_{c}-\varepsilon^{j}_{p_2}
)\\&\times[(f^{j}_{2}+1)f^{k}_{1}-f^{j}_{2}(f^{k}_{1}+1)]\delta({\bf p}-{\bf p}_1).
\label{c12jk2}
\end{align}
{Although 
physically intuitive,} this term is qualitatively different to the 
collision integral of Eq.~\eqref{c12jk}, as it describes a process (Fig.~\ref{fig2}(iii)) whereby one condensate and one non-condensate atom from different components scatter into a thermal and condensed state respectively. 

The final collision processes $C^{jj}_{22}$ and $C^{kj}_{22}$ in Eq.~(\ref{kin}) describe the scattering  between non-condensate atoms of 
same ($k=j$) or differing ($k \neq j$) species, given by
\begin{align}\nonumber
&C^{kj}_{22}=(1+\delta_{kj})\frac{g_{kj}^2}{(2\pi)^5\hbar^7}\int d{\bf p}_2\int d{\bf p}_3\int d{\bf p}_4\\\nonumber&\times\delta({\bf p}+{\bf p}_2-{\bf p}_3-{\bf p}_4)\delta(\varepsilon^{j}_{p}+\varepsilon^{k}_{p_2}-\varepsilon^{k}_{p_3}-\varepsilon^{j}_{p_4})\\&[(f^{j}+1)(f^{k}_{2}+1)f^{k}_{3}f^{j}_{4}-f^{j}f^{k}_{2}(f^{k}_{3}+1)(f^{j}_{4}+1)].
\label{c22jk}
\end{align}


{\it Numerical Results. } 
{To gain insight into the relative importance of the collision rates 
we 
compute their temperature dependence 
for experimentally-relevant equilibrium $^{87}$Rb-$^{41}$K and $^{87}$Rb-$^{85}$Rb mixtures} in isotropic harmonic traps ($\omega=2\pi\times20$Hz), and a total atom number $N_{j}=10^{5}$ in each component. These mixtures were chosen as their tunable scattering lengths \cite{modugno_2002,thalhammer_2008,papp_2008} 
enable the probing of both miscible $\Lambda=g_{12}/\sqrt{g_{11}g_{22}}<1$ and immiscible ($\Lambda > 1$) regimes. By re-expressing $C_{12}^{kj} = C_{12}^{kj, {\rm out}} - C_{12}^{kj, {\rm in}}$ (and analogously for $\mathds{C}_{12}^{kj}$ and $C^{kj}_{22}$), we explicitly identify "in" and "out" scattering rates, which are equal at equilibrium. Following Refs.~\cite{jackson_2002b,zaremba_1999}, we define collisional rates $\Gamma_{12(22)}^{kj}=\int {d\vec{p}}/(2\pi\hbar)^3C_{12(22)}^{kj,{\rm out}}$ that give the number of atoms leaving a phase-space volume $d\vec{r}d\vec{p}/h^3$ per unit time as a result of collisions, for a perturbation from equilibrium.
%
%
By transforming to the centre of mass frames, the collision rates can be written as
$\Gamma^{kj}_{12}=\int\frac{d{\bf p}_2}{(2\pi\hbar)^3}f^{k}_{2}\,n_{c,j}\,\sigma_{kj}v_{r}\int\frac{d\Omega}{4\pi}(f^{k}_{3}+1)(f^{j}_{4}+1),$ where $v_{r}$
is the relative velocity and 
$\sigma_{kj}=(1+\delta_{kj})4\pi a_{kj}^{2}$ the cross-section. 
The collision rates between non-condensate atoms are $\Gamma^{kj}_{22}=\int\frac{d{\bf p}_1}{(2\pi\hbar)^3}f^{j}_{1}\int\frac{d{\bf p}_2}{(2\pi\hbar)^3}f^{k}_{2}\int\frac{d\Omega}{4\pi}\sigma_{kj}\allowbreak|{\bf v}_1-{\bf v}_2|(f^{k}_{3}+1)(f^{j}_{4}+1),\label{g22}$
while the $\mathds{C}^{kj}_{12}$ collisions scattering rate, 
$\Gamma^{kj}_{\mathds{C}}=\int {d\vec{p}}/{(2\pi\hbar)^3}\mathds{C}_{12}^{kj,{\rm out}}$, takes the form
%
\begin{equation}
\label{gc}
\Gamma^{kj}_{\mathds{C}}=\sigma_{kj}\left(\frac{\mathcal{M}_{kj}}{m_{kj}}\right)^2n_{c,k}\,n_{c,j}\,\tilde{v}_{r}\int\frac{d\Omega}{4\pi}f^{j}_{2}(f^{k}_{1}+1),
\end{equation}
with $\tilde{v}_{r}$
 the relative velocity and $\mathcal{M}_{kj}^{-1}=m_{k}^{-1}-m_{j}^{-1}$. 
%
%

\begin{figure}[t]
\includegraphics[scale=0.7]{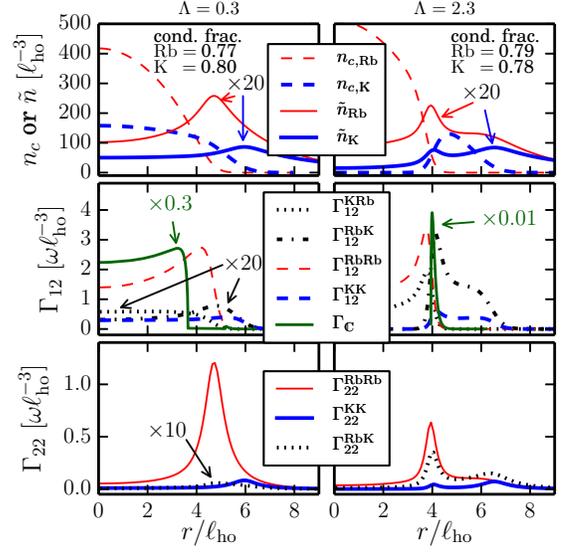}\caption{\label{fig:RbK_T21}(Color online) Miscible (left) and immiscible (right) $^{87}$Rb-$^{41}$K mixture in an isotropic harmonic trap (trap frequency $\omega=2\pi\times20$Hz) at temperature $21$nK with scattering lengths $a_{\rm Rb87}=99a_0$, $a_{\rm K}=60a_0$, $a_{\rm Rb-K}=20a_0$(miscible) or $163a_0$(immiscible) \cite{modugno_2002,thalhammer_2008}; 
each species has a total of $N=10^5$ atoms. 
{While our model does not include critical fluctuations required for an accurate determination of the critical temperature, an estimate for this can be obtained
from numerical fits of our self-consistent condensate fractions by $1-(T/T_c)^\alpha$, where $T_c$ and $\alpha$ are fitting parameters; we find $T_c\approx39$nK, which is lower than the mean-field single-component $T_c$~\cite{dalfovo_1999} by at most 5\%.} (Top) Condensate and thermal densities. (Middle) Spatially-resolved collision rates between condensate and thermal atoms.  (Bottom) Spatially-resolved collision rates between thermal atoms.}
\end{figure}

Figure \ref{fig:RbK_T21} shows the equilibrium condensate/thermal density profiles (top panels) and collision rates (middle/bottom panels) for a mixture of $^{87}$Rb and $^{41}$K at a temperature of $T=21{\rm nK}$, when condensate fractions $\approx 80\%$, for both miscible ($\Lambda=0.3$, left column) and immiscible ($\Lambda=2.3$, right column) cases. 
The two condensates (dashed lines) mix (top panel, left) or phase-separate (top panel, right) with thermal clouds (solid lines) displaying peaks at the condensate edges, thus also leading to a mean-field-induced double-peaked thermal structure.
The spatially-resolved collisional rates between condensate and non-condensate (middle panels) reveal peaks close to the condensate edges, which become more pronounced for immiscible condensates.
In particular, $\Gamma_{\mathds{C}}^{kj}$ collisional rates feature large localized peaks in regions where both components exhibit an appreciable condensate, which can locally dominate all other collisional processes in the centre of the immiscibility region.
Collision rates between non-condensate atoms (bottom panels) are found to closely follow the shape of the thermal 
profiles, with the inter-species rates, 
$\Gamma^{\text{RbK}}_{22}$ (black-dotted line), affected by both thermal cloud distributions.

\begin{figure}[t!]
\includegraphics[scale=0.75]{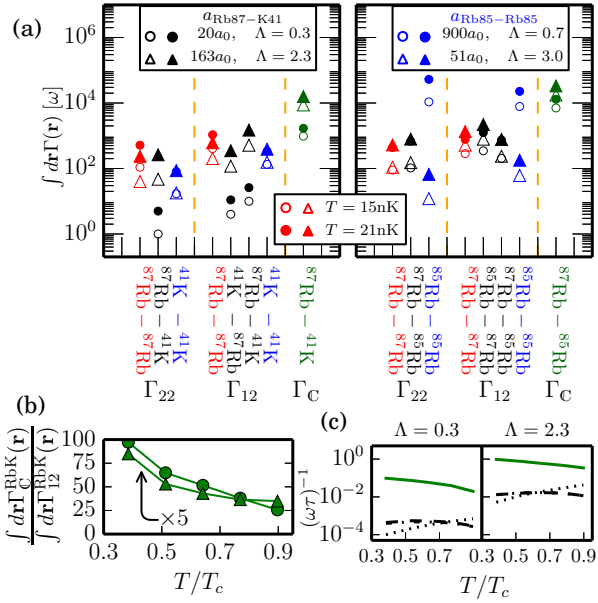}\caption{(Color online) (a) Comparison of integrated collision rates among the various processes at different temperatures and miscibilities for (left) $^{87}$Rb-$^{41}$K  ($a_{\rm Rb}=99a_0$, $a_{\rm K}=60a_0$) and (right) $^{87}$Rb-$^{85}$Rb ($a_{\rm Rb87-Rb87}=99a_0$, $a_{\rm Rb87-Rb85}=213a_0$~\cite{papp_2008}); 
other parameters as in Fig.~\ref{fig:RbK_T21}. (Bottom) Corresponding temperature dependence for $^{87}$Rb-$^{41}$K mixtures of (b) ratio of integrated collision rates of $\mathds{C}^{\rm RbK}_{12}$ to $C^{\rm RbK}_{12}$, and {(c) collision times $(\omega\tau)^{-1}$ of $C_{22}^{\rm RbK}$(dots), $C_{12}^{\rm RbK}$ and $C_{12}^{\rm KRb}$ (dashes/dashed-dots) and $\mathds{C}^{\rm RbK}_{12}$ (solid) collisions.}}
\label{fig:rates}
\end{figure}

The cumulative effect of such collisional processes is best characterized through the
%
%
integrated collision rates 
$\int d {\bf r} \Gamma({\bf r})$ of
$\Gamma^{kj}_{22}$, $\Gamma^{kj}_{12}$ and $\Gamma^{kj}_{\mathds{C}}$ over the cloud's volume. 
These are shown in Fig.~\ref{fig:rates}(a) for $^{87}$Rb-$^{41}$K (left panels), and $^{87}$Rb-$^{85}$Rb mixtures (right panels), in their experimentally-accessible
miscibility (circles) and immiscibility (triangles) regimes for $T=21$nK (Fig.~3 data, filled symbols) and $T=15$nK (open symbols),. We find: 
%
(i) the 
${\mathds{C}_{12}^{kj}}$ collisions (green symbols) are at least as large as the other collisional terms in all cases; 
(ii) increasing the temperature (hollow to filled symbols) in the presence of an appreciable condensate fraction, enhances all collisional rates involving thermal atoms (including $C_{12}^{kj}$ and ${\mathds{C}_{12}^{kj}}$); 
(iii) collisional rates are largely sensitive to the interation strength $g_{kj}$ (through $g_{kj}^2$ prefactors), controllable through Feshbach resonances: 
an increase in the inter-species scattering length $a_{\text{Rb87-K41}}$~\cite{modugno_2002,thalhammer_2008} (left image) increases the relative importance of all $^{87}$Rb-$^{41}$K
thermal-thermal, or condensate-thermal
 collisional terms (black/green points).
Likewise, an increase in the tunable intra-component scattering length $a_{\text{Rb85-Rb85}}$~\cite{papp_2008} (right), 
enhances the $^{85}$Rb-$^{85}$Rb (blue) collisional terms.
%

{The significance of the ${\mathds{C}_{12}^{kj}}$ collisional process becomes evident in Fig.~\ref{fig:rates}(b), showing the temperature dependence of its ratio to the corresponding integrated $\Gamma_{12}^{\rm RbK}$ rates for $^{87}$Rb-$^{41}$K mixtures.
Since the efficiency of sympathetic cooling depends on the energy exchange rate, the rate of interspecies collisions~\cite{delannoy_2001} demonstrates the relevant role that the ${\mathds{C}_{12}^{kj}}$ collisional terms may play 
\begin{table}[H]
\centering
\resizebox{0.35\textwidth}{!} {
{\begin{tabular}{|c || c | c | c | c | c |c | }\hline
   \multirow{3}{*}{}   & \multicolumn{3}{c|}{$\Lambda=0.7$} & \multicolumn{3}{c|}{$\Lambda=3.0$} \\\cline{2-7}
   & { $^{87}$Rb}& {$^{87}$Rb} & {$^{85}$Rb} & {$^{87}$Rb} & { $^{ 87}$Rb} & { $^{85}$Rb} \\
   & {$^{85}$Rb} & {$^{87}$Rb} & {$^{85}$Rb} &  { $^{ 85}$Rb} & { $^{ 87}$Rb} & { $^{ 85}$Rb}\\\hline   
$(\omega\tau_{22})^{-1}$ & $ 10^{ -1}$ & $ 10^{ -2}$&1& $ 10^{ -1}$& $ 10^{ -2}$& $ 10^{ -2}$\\\hline
$(\omega\tau_{12})^{-1}$&  $ 10^{ -2}$ &  $ 10^{ -2}$ &  1 &  $ 10^{ -1}$ &  $ 10^{ -1}$ &  $ 10^{ -2}$\\\hline
$(\omega\tau_{\mathds{C}})^{-1}$ & 1&-&-&1&-&- \\\hline
\end{tabular}}}\vspace{0.3cm}
\caption{{\label{tab1}Hydrodynamic parameters $(\omega\tau)^{-1}$ (in order-of-magnitude) for the $^{87}$Rb-$^{85}$Rb mixtures.}}
\end{table}
\noindent during various sympathetic cooling stages.} 

{The evaluation of integrated collision rates enables us to define a typical collisional timescale $\tau$, as $\tau^{-1} = \int d\vec{r} \Gamma(\vec{r})/N_{\rm coll}$, where $N_{\rm coll}$ is the relevant number of available thermal atoms taking part in collisions for each process; 
this allows us to
distinguish between the collisionless [$(\omega\tau)^{-1}<1$] and the hydrodynamic [$(\omega\tau)^{-1}>1$] regimes~\cite{nikuni_1999,pethick_2008}.
%
Fig.~\ref{fig:rates}(c) shows the variation of the hydrodynamic parameter $(\omega\tau)^{-1}$ for $^{87}$Rb-$^{41}$K mixtures 
in the miscible (left) and immiscible (right) regimes, focussing only on interspecies collisions: this reveals the dominance of $(\omega\tau_{\mathds{C}})^{-1}$, which approaches 1 in the immiscible case, with all intra-species collisions satisfying 
$10^{-2}<(\omega\tau)^{-1}<10^{-1}$ in both cases. 
Doubling the trap frequency to $2 \pi \times 40$Hz roughly doubles $(\omega\tau_{22})^{-1}$, increases $(\omega\tau_{12})^{-1}$ by $\approx 50\%$ and only slightly increases $(\omega\tau_{\mathds{C}})^{-1}$, which however remains the largest.
Consideration of realistic $^{87}$Rb-$^{85}$Rb mixtures [characteristic values shown in Table \ref{tab1}] reveals that while $(\omega\tau_{\mathds{C}})^{-1} \sim 1$ is always the largest interspecies contribution, tuning the intraspecies $a_{\text{Rb85-Rb85}}$ to 900$a_0$~\cite{papp_2008} provides a plausible candidate for multi-component hydrodynamic behaviour in the sense of
$(\omega\tau_{\mathds{C}})_{\rm inter}^{-1} \sim (\omega\tau_{12})_{\rm intra}^{-1} \sim (\omega\tau_{22})_{\rm intra}^{-1} \sim 1$.}

{\it Conclusions. }
We have presented a system of coupled kinetic equations for the evolution of two-component condensates in the presence of dynamical thermal clouds,. 
{Analytical considerations led to the identification of a "cross-condensate-exchange" term, which was numerically shown to dominate, even at relatively high temperatures, over other collisional terms close to equilibrium, both for $^{87}$Rb-$^{41}$K and $^{87}$Rb-$^{85}$Rb mixtures. 
Such a term could be highly relevant during sympathetic cooling, particularly in later stages, when the sympathetically-cooled component acquires appreciable condensation.
%
%
Consideration of collisional timescales indicated the potential of generating hydrodynamic multi-component condensates with $(\omega \tau)^{-1} \sim 1$, at least for some collisional processes.}
Our work, which is generalizable to multi-component Bose gases and Bose-Fermi mixtures, sets the scene for future dynamical studies of sympathetic cooling, multi-component condensate formation {and coupled expansion dynamics.} 

{\it Acknowledgements. } We acknowledge discussions with S.L. Cornish, S.A. Gardiner, P. Mason, E. Zaremba and M. Edwards, and support from EPSRC (Doctoral Prize Fellowship (MJE), grant EP/K03250X/1 (KLL, NPP)).

\end{document}